\documentclass[runningheads]{llncs}
\usepackage{graphicx,color}
\newcommand{\comentario}[1]{}  %comment not showed
\newcommand{\respuesta}[1]{}  %respuesta not showed
\usepackage{savesym}
\usepackage{booktabs}
\usepackage{multirow}
\usepackage{amsmath}
\usepackage{caption}
\usepackage{subcaption}
\usepackage{tabularx}
\makeatletter
\newcommand{\@chapapp}{\relax}%
\makeatother
\usepackage[title]{appendix}
\usepackage{pdflscape}

\begin{document}
\title{Vocabulary-based Method for Quantifying Controversy in Social Media}
%
%\titlerunning{Abbreviated paper title}
% If the paper title is too long for the running head, you can set
% an abbreviated paper title here
%
%\author{Juan Manuel Ortiz de Zarate \and Esteban Feuerstein}
%
\author{Juan Manuel Ortiz de Zarate \and Esteban Feuerstein}
%
%\author{Juan Manuel Ortiz de Zarate and Esteban Feuerstein}
% First names are abbreviated in the running head.
% If there are more than two authors, 'et al.' is used.
%
\institute{Universidad de Buenos Aires}
%\author{} %\and
%Second Author\inst{2,3}\orcidID{1111-2222-3333-4444} \and
%Third Author\inst{3}\orcidID{2222--3333-4444-5555}}
%
\authorrunning{Ortiz de Zarate et al.}

%Springer Heidelberg, Tiergartenstr. 17, 69121 Heidelberg, Germany
%\email{lncs@springer.com}\\
%\url{http://www.springer.com/gp/computer-science/lncs} \and
%ABC Institute, Rupert-Karls-University Heidelberg, Heidelberg, Germany\\
%\email{\{abc,lncs\}@uni-heidelberg.de}}
%
\maketitle              % typeset the header of the contribution
\begin{abstract}
Identifying controversial topics is not only interesting from a social point of view, it also enables the application of methods to avoid the information segregation, creating better discussion contexts and reaching agreements in the best cases. 
In this paper we develop a systematic method for controversy detection based primarily on the jargon used by the communities in social media. 
Our method dispenses with the use of domain-specific knowledge, is language-agnostic, efficient and easy to apply. We perform an extensive set of experiments across many languages, regions and contexts, taking controversial and non-controversial topics. We find that our vocabulary-based measure performs better than state of the art measures that are based only on the community graph structure. Moreover, we shows that it is possible to detect polarization through text analysis.
%\keywords{Social Networks \and NLP \and Controversy}
\end{abstract}
\section{Introduction}

\label{sec:introduction}
Controversy is a phenomenom with a high impact at various levels. It has been broadly studied from the perspective of different disciplines, ranging from the seminal analysis of the conflicts within the members of a karate club \cite{zachary1977information} to political issues in modern times \cite{conover2011political,morales2015measuring}. The irruption of digital social networks \cite{easley2010networks} gave raise to new ways of intentionally intervening on them for taking some advantage \cite{stewart2018examining,calvo2015anatomia}. Moreover highly contrasting points of view in some groups  tend to provoke conflicts that lead to attacks from one community to the other by harassing, “brigading”, or “trolling” it  \cite{kumar2018community}.
The existing literature shows different issues that controversy brings up such as splitting of communities, biased information, hateful discussions and attacks between groups,
generally proposing ways to solve them. For example, Kumar, Srijan, et al. \cite{kumar2018community} analyze many techniques to defend us from attacks in \textit{Reddit} \footnote{https://www.reddit.com/} while Stewart, et al. \cite{stewart2018examining} insinuate that there was external interference in Twitter during the 2016 US presidential elections to benefit one candidate. 
%\comentario{por qué usás la doble barra para punto y aparte? No funciona el clásico doble enter?}
Also, as shown in~\cite{kulshrestha2015characterizing}, detecting controversy could provide the basis to improve the \textit{``news diet"} of readers, offering the possibility to connect users with different points of views by recommending them new content to read \cite{munson2013encouraging}.

Moreover, other studies on “bridging echo chambers” \cite{garimella2017reducing} and the positive effects of intergroup dialogue \cite{allport1954nature,pettigrew2013does} suggest that direct engagement could be effective for mitigating such conflicts. 
Therefore,  easily and automatically identifying controversial topics could allow us to quickly implement  different strategies for preventing miss-information, fights and bias. \textit{Quantifying} the controversy is even more powerful, as it allows us to establish controversy levels, and in particular to classify controversial and non-controversial topics by establishing a threshold score that separates the two types of topics. With this aim, we propose in this work a systematic, language-agnostic method to quantify controversy on social networks taking tweet's content as root input. Our main contribution is a new  vocabulary-based method  that works in any language and equates the performance of state-of-the-art structure-based methods. Finally, controversy quantification through vocabulary analysis opens several research avenues to analyze whether polarization is being created, maintained or augmented  by the ways of talking of each community. 

Having this in mind and if we draw from the premise that when a discussion has a high controversy it is in general due to the presence of two principal communities fighting each other (or, conversely, that when there is no controversy there is just one principal community the members of which share a common point of view), we can measure the controversy by detecting if the discussion has one or two principal jargons in use. 
Our method is tested on Twitter datasets. This microblogging platform has been widely used to analyze discussions and polarization \cite{rajadesingan2014identifying,yardi2010dynamic,trilling2015two,weller2014twitter,morales2015measuring}. It is a natural choice for these kind of problems, as it represents one of the main fora for public debate in online social media \cite{weller2014twitter}, it is a common destination for affiliative expressions \cite{hong2012online} and is often used to report and read news about current events \cite{shearer2017news}. An extra advantage of Twitter for this kind of studies is the availability of real-time data generated by millions of users. Other social media platforms offer similar data-sharing services, but few can match the amount of data and the accompanied documentation provided by Twitter. One last asset of Twitter for our work is given by \emph{retweets}, whom typically indicate endorsement \cite{bessi2014social} and hence become a useful concept to model discussions as we can set  ``who is with who". However, our method has a general approach and it could be used a priori in any social network. In this work we report excellent result tested on Twitter but in future work we are going to test it in other social networks.

Our paper is organized as follows: in Section \ref{sec:related_work}, we review related work. Section \ref{sec:method} contains the detailed explanation of the pipeline we use for quantifying controversy of a topic, and each of its stages. In Section \ref{sec:experiments} we report the results of an extensive empirical evaluation of the proposed measure of controversy. Finally, Section \ref{sec:discussion} is devoted to discuss possible improvements and directions for future work, as well as lessons learned.

\section{Related work}
\label{sec:related_work}

Many previous works are dedicated to quantifying the polarization observed in online social networks and social media \cite{conover2011political,guerra2013measure,amelkin2017distance,akoglu2014quantifying,dandekar2013biased,garimella2018quantifying}. The main characteristic of those works is that the measures proposed are based on the structural characteristics of the underlying graph. Among them, we highlight the work of Garimella et al.\cite{garimella2018quantifying} that presents an extensive comparison of controversy measures, different graph-building approaches, and data sources, achieving the best performance of all.
%\comentario{no entendí qué hacen estos. Si son anteriores a Garimella quizás habría que arrancar con estos, y decir algo así como ``Hay muchos trabajos previos que cuantifican basándose en la estructura del grafo, por ejemplo blbablabla que tiene estas características: blabla. Entre todos estos destacamos a Garmiella...." y ahí arrancar con lo de Garimella}
In their research they propose different metrics to measure polarization on Twitter.  Their techniques based on the structure of the endorsement graph can successfully detect whether a discussion (represented by a set of tweets), is controversial or not regardless of the context and most importantly, without the need of any domain expertise. They also consider two different methods to measure controversy based on the analysis of the posts contents, but both fail when used to create a measure of controversy. 

Matakos et al. \cite{matakos2017measuring} develop a \textit{polarization index}. Their measure captures the tendency of opinions to concentrate in network communities, creating echo-chambers. They obtain a good performance at identifying controversy by taking into account both the network structure and the existing opinions of users. However, they model opinions as positive or negative with a real number between -1 and 1. Their performance is good, but although it is an opinion-based method it is not a text-related one.Other recent works \cite{ramponi2019vocabulary,tran2016characterizing,lahoti2018joint} have shown that communities may express themselves with different terms or ways of speaking, use different jargon, which in turn can be detected with the use of text-related techniques. 

 In his thesis \cite{jang2019probabilistic}, Jang explains controversy via generating a summary of two conflicting stances that make up the controversy. This work shows that a specific sub-set of tweets could represent the two opposite positions in a polarized debate.

A good tool to \emph{see} how communities interact is ForceAtlas2~\cite{jacomy2014forceatlas2}, a force-directed layout widely used for visualization. This layout has been recently found to be very useful at visualizing community interactions \cite{venturini2019we}, as this algorithm will draw groups with little communication between them in different areas, whereas, if they have many interactions they will be drawn closer to each other. Therefore, whenever there is controversy the layout will show two  well separated groups and will tend to show only one big community otherwise.

The method we propose to measure the controversy equates in accuracy
the one developed by Garimella et al.\cite{garimella2018quantifying} and improves considerably computing time and robustness wrt the amount of data needed to effectively apply it. Our method is also based on a graph approach but it has its main focus on the vocabulary. We first  train an NLP classifier that estimates opinion polarity of main users, then we run label-propagation~\cite{zhur2002learning} on the endorsement graph to get polarity of the whole network. 
%\comentario{to get polarity of the whole users no está bien} 
Finally we compute the controversy score through a computation inspired in Dipole Moment, a measure used in physics to estimate electric polarity on a system.
%\comentario{te robé la descripción del algoritmo y la pasé a la sección siguiente, antes del paso a paso. Completala acá con el mismo nivel de detalle que los related works. O sea, lo que te robé pero sin mencionar fasttext, etc.Enfatizar que nuestro método combina aspectos de lenguaje con aspectos de red}
In our experiments we use the same data-sets from other works ~\cite{darwish2019quantifying,garimella2018quantifying,garimella2016exploring} as well as other datasets that we collected by us using a similar criterion (described in Section \ref{sec:experiments}).

\section{Method}
\label{sec:method}
Our approach to measuring controversy is based on a systematic way of characterizing social media activity through its content. We employ a pipeline with five stages, namely \textit{graph building}, \textit{community identification}, \textit{model training}, \textit{predicting} and \textit{controversy measure}. The final output of the pipeline is a value that measures how controversial a topic is, with higher values corresponding to higher degrees of controversy. The method is based on analysing posts content  through Fasttext~\cite{joulin2016bag}, a library for efficient learning of word representations and sentence classification developed by Facebook Research team.
In short, our method works as follows: through Fasttext we train a language-agnostic model which can predict the community of many users by their jargon. Then we take there predictions and compute a score based on the physic notion \textit{Dipole Moment} \footnote{In physics, the electric dipole moment is a measure of the separation of positive and negative electrical charges within a system, that is, a measure of the system's overall polarity}using a language approach to identify core or characteristic users and set the polarity trough them. We provide a detailed description of each stage in the following.

\vspace{5pt}
\textbf{Graph Building}

This paragraph provides details about the approach used to build graphs from raw data. As we said in Section \ref{sec:introduction}, we  extract our discussions from Twitter. Our purpose is to build a conversation graph that represents activity related to a single topic of discussion -a debate about a specific event. 

For each topic, we build a graph $G$ where we assign a vertex to each user who contributes to it and we add a directed edge from node $u$ to node $v$ whenever user $u$ retweets a tweet posted by $v$.
Retweets typically indicate endorsement \cite{bessi2014social}: users who retweet signal endorsement of the opinion expressed in the original tweet by propagating it further. Retweets are not constrained to occur only between users who are connected in Twitter's social network, but users are allowed to retweet posts generated by any other user. As many other works in literature \cite{calvo2015anatomia,bild2015aggregate,kupavskii2012prediction,feng2013retweet,stewart2018examining,morales2015measuring} we establish that one retweet among a pair of users are needed to define an edge between them.

\vspace{5pt}
\textbf{Community Identification}

To identify a community's jargon we need to be very accurate at defining its members. If we, in our will of finding two principal communities,  force the partition of the graph in that precise number of communities, we may be adding noise in the jargon of the principal communities that are fighting each other. 
Because of that, we decide to cluster the graph trying two popular algorithms: Walktrap \cite{pons2005computing} and Louvain \cite{blondel2008fast}. Both are structure-based algorithms that have very good performance with respect to the Modularity Q measure\footnote{Q(G)=$\sum_{C \in G}(e_{c}-a_{c})$, where $G$ is the graph, $C$ each of its communities, $e_{c}$ the fraction of internal edges and $a_{c}$ the fraction of edges in the border}. These techniques does not detect a fixed number of clusters; their output will depend on the  Modularity Q optimization, resulting in less ``noisy" communities. The main differences between the two methods, in what regards our work, are that Louvain is a much faster heuristic algorithm but produces clusters with worse Modularity Q. Therefore, in order to analyze the trade-off between computing time and quality we decide to test both methods.
At this step we want to capture the tweets of the principal communities to create the model that could differentiate them. Therefore, we take the two communities identified by the cluster algorithm that have the maximum number of users, and use them for the following step of our method.

\vspace{5pt}
\textbf{Model Training}

After detecting the principal communities we create our training dataset to feed the model. To do that, we extract the tweets of each cluster, we sanitize and we subject them to some transformations. First, we remove duplicate tweets -e.g. retweets without additional text. Second, we remove from the text of the tweets user names, links, punctuation, tabs, leading and lagging blanks, general spaces and ``RT" - the text that points that a tweet is in fact a retweet. \\
As shown in previous works, emojis\footnote{https://emojipedia.org/twitter/}  
are correlated with sentiment \cite{novak2015sentiment}. Moreover, as we think that communities will express different sentiment during discussion, it is forseeable that emojis will play an important role as separators of tweets that differentiate between the two sides. Accordingly, we decide to add them to the train-set by translating each emoji into a different word. For example, the emoji :) will be translated into \textit{happy} and :( into \textit{sad}. Relations between emojis and words are defined in the R library \textit{textclean}\footnote{https://cran.r-project.org/web/packages/textclean/textclean.pdf}.

Finally, we group tweets by user concatenating them in one string
and labeling them with the user's community, namely with tags \textit{C1} and \textit{C2}, corresponding respectively to the biggest and  second biggest groups. 
It is important to note that we take the same number of users of each community to prevent bias in the model. Thus, we use the number of users of the smallest principal community. 

The train-set built that way is used to feed the model. As we said, we use Fasttext \cite{joulin2016bag} to do this training. To define the values of the hyper-parameters we use the findings of~\cite{yang2018using}. In their work they investigate the best hyper-parameters to train word embedding models using Fasttext \cite{joulin2016bag} and Twitter data. We also change the default value of the hyper-parameter \textit{epoch} to 20 instead of 5 because we want more convergence preventing as much as possible the variance between different training. These values could change in other context or social networks where we have more text per user or different discussion dynamics.

\vspace{5pt}
\textbf{Predicting}

The next stage consists of identifying the \textit{characteristic users} of each side the discussion. These are the users that better represent the jargon of each side. To do that, tweets of the users belonging to the largest connected component of the graph are sanitized and transformed exactly as in the Training step.

We decide to restrict to the largest connected component because in all cases it contains more than 90\% of the nodes. The remaining 10\% of the users don't participate in the discussion from a collective point of view but rather in an isolated way and this kind of intervention does not add interesting information to our approach. Then, we remove from this component users with degree smaller or equal to 2 (i.e. users that were retweeted by another user or retweeted other person less than three times in total). Their  participation in the discussion is marginal, consequently they are not relevant wrt controversy as they add more noise than information at measuring time. This step could be adjusted differently in a different social network. We name this result component \textit{root-graph}.

Finally, let's see how we do classification. Considering that Fasttext returns for each classification both the predicted tag and the probability of the prediction, we classify each user of the resulting component by his sanitized tweets with our trained model, and take users that were tagged with a probability greater or equal than 0.9. These are the \textit{characteristic users} that will be used in next step to compute the controversy measure.

\vspace{5pt}
\textbf{Controversy Measure}
%\vspace{-1pt}

This section describes the controversy measures used in this work. This computation 
is inspired in the measure presented by Morales et al. \cite{morales2015measuring}, and is based on the notion of dipole moment that has its origin in physics. 

First, we assign to the \textit{characteristic users} the probability returned by the model, negativizing them if the predicted tag  was \textit{C2}. Therefore, these users are  assigned values in the set [-1,-0.9] $\cup$ [0.9,1]. Then, we set values for the rest of the users of the \textit{root-graph} by label-propagation \cite{zhur2002learning} - an iterative algorithm to propagate values through a graph by node's neighborhood.

Let $n^{+}$ and $n^{-}$ be the number of vertices $V$ with positive and negative values, respectively, and $\Delta A = \dfrac{\mid n^{+} - n^{-}\mid}{\mid V \mid}$ the absolute difference of their normalized size. Moreover, let $gc^{+}$ ($gc^{-}$) be the average value among vertices $n^{+}$ ($n^{-}$) and set $\tau$ as half their absolute difference, $\tau = \dfrac{\mid gc^{+} - gc^{- }\mid}{2}$. The dipole moment content controversy measure is defined as: $\textit{DMC} = (1 -\Delta A)\tau$.

The rationale for this measure is that if the two sides are well separated, then label propagation will assign different extreme values to the two partitions, where users from one community will have values near to 1 and users from the other to -1, leading to higher values of the \textit{DMC} measure. Note also that larger differences in the size of the two partitions (reflected in the value of $\Delta A$) lead to smaller values for the measure, which takes values between zero and one.

\section{Experiments}
\label{sec:experiments}
In this section we report the results obtained by running the above proposed method over different discussions. 
\subsection{Topic definition}
In the literature, a topic is often defined by a single hashtag. However, this might be too restrictive in many cases. In our approach, a topic is operationalized as an specific hashtags or \textit{key words}. Sometimes a discussion in a particular moment could not have a defined hashtag but it could be around a certain \textit{keyword}, i.e. a word or expression that is not specifically a hashtag but it is widely used in the topic. For example during the Brazilian presidential elections in 2018 we captured the discussion by the mentions to the word \textit{Bolsonaro}, that is the principal candidate's surname.\\
Thus, for each topic we retrieve all the tweets that contain one of its hashtags or the \textit{keyword} and that are generated during the observation window. We also ensure that the selected topic is associated with a large enough volume of activity.
\subsection{Datasets}
In this section we detail the discussions we use to test our metric and how we determine the ground truth (i.e. if the discussion is controversial or not).
We use thirty different discussions that took place between March 2015 and June 2019, half of them with controversy and half without it. We considered discussions in four different languages: English, Portuguese, Spanish and French, occurring in five regions over the world: South and North America, Western Europe, Central and Southern Asia. We also studied these discussions taking first 140 characters and then 280 from each tweet to analyze the difference in performance and computing time wrt the length of the posts.

To define the amount of data needed to run our method we established that the Fasttext model has to predict at least one user of each community with a probability greater or equal than 0.9 during ten different trainings. If that is not the case, we are not able to use \textit{DPC} method. This decision made us consider only a subset of the datasets used in \cite{garimella2018quantifying}, because due to the time elapsed since their work, many tweets had been deleted and consequently the volume of the data was not enough for our framework. To enlarge our experiment base we added new debates, more detailed information about each one is shown in Table~\ref{tab:datasets} in~\ref{apendice}. To select new discussions and to determine if they are controversial or not we looked for topics widely covered by mainstream media, and that have generated ample discussion, both online and offline. For non-controversy discussions we focused on ``soft news" and entertainment, but also to events that, while being impactful and/or dramatic, did not generate large controversies. To validate that intuition, we manually checked a sample of tweets, being unable to identify any clear instance of controversy\footnote{Code and networks used in this work are available - Omitted for anonymity reasons} %\url{http://github.com/jmanuoz/Vocabulary-based-Method-for-Quantify-Controversy}}. 
On the other side, for controversial debates we focused on political events such as elections, corruption cases or justice decisions. 

\begin{figure}
\centering
\begin{subfigure}[b]{0.45\textwidth}
\centering
\includegraphics[width=\textwidth]{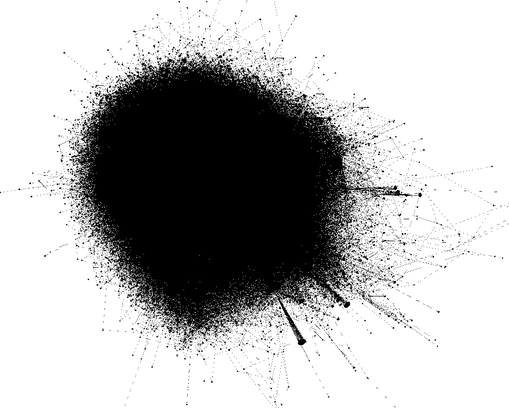}
\caption{ForceAtlas2 layout over the root graph of Halsey discussion}
\label{fig:halsey}
\end{subfigure}
\begin{subfigure}[b]{0.45\textwidth}
\centering
\includegraphics[width=\textwidth]{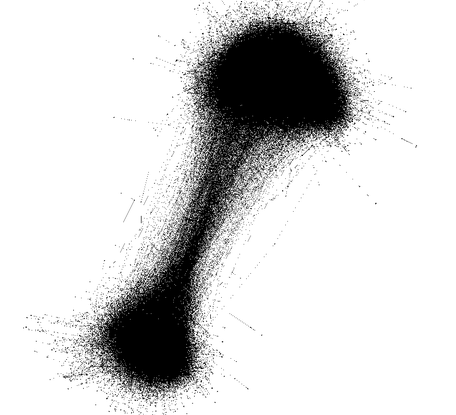}
\caption{ForceAtlas2 layout over the root graph of Kavanaugh discussion}
\label{fig:kavanaugh}
\end{subfigure}
\caption{}
\end{figure}

To furtherly establish the presence of absence of controversy of our datasets, we visualized the corresponding networks through ForceAtlas2~\cite{jacomy2014forceatlas2}. Figures \ref{fig:halsey} and \ref{fig:kavanaugh} show an example of how non-controversial and controversial discussions look like respectively with ForceAtlas2 layout.  As we can see in these figures, in a controversial discussion this layout tends to show two well separated groups while in a non-controversial one it tends to be only one big group. More information on the discussions is given in Table~\ref{tab:datasets}.

To avoid potential overfitting, we use only twelve graphs as testbed during the development of the measures, half of them controversial (netanyahu, ukraine, @mauriciomacri 1-11 Jan, Kavanaugh 3 Oct, @mauriciomacri 11-18 Mar, Bolsonaro 27 Oct) and half non-controversial (sxsw, germanwings, onedirection, ultralive, nepal, mothersday). This procedure resembles a 40/60\% train/test split in traditional machine learning applications.

Some of the discussions we consider refer to the same topics but in different periods of time. We needed to split them because our computing infrastructure does not allow us to compute such an enormous amount of data. However, being able to estimate controversy with only a subset of the discussion is an advantage, because discussions could take many days or months and we want to identify controversy as soon as possible, without the need of downloading the whole discussion. Moreover, for very long lasting discussions in social networks gathering the whole data would be impractical for any method.

\vspace{-15pt}
\subsection{Results}

\begin{landscape}
\begin{figure}[htbp]

\begin{subfigure}[t]{0.78\textheight}
    \includegraphics[width=\linewidth,height=5cm]{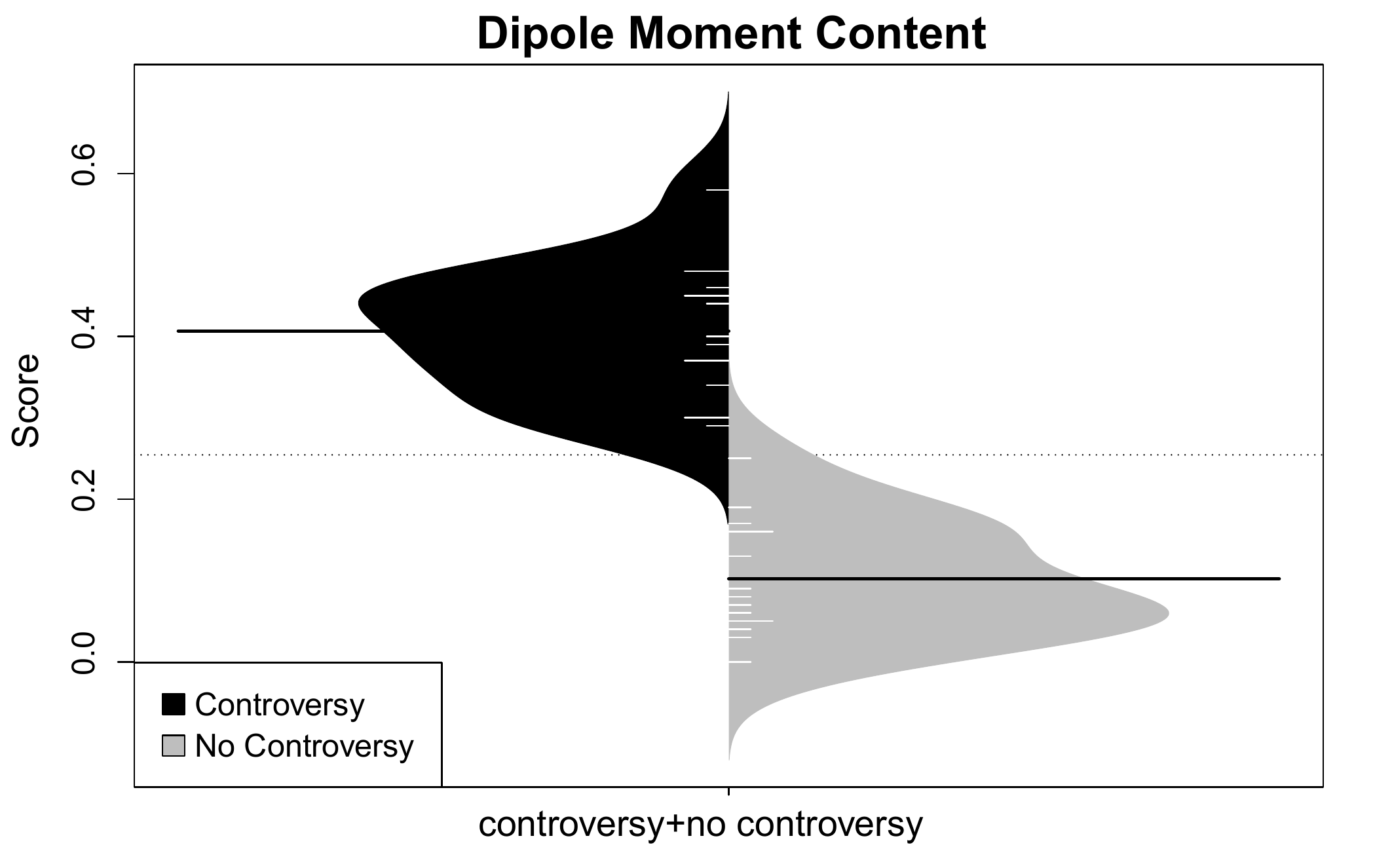}
\caption{Average controversy scores over 20 runs on datasets with 280 characters}
\label{fig:scores_prom}
\end{subfigure}\hfill
\begin{subfigure}[t]{0.78\textheight}
    \includegraphics[width=\textwidth,height=5cm]{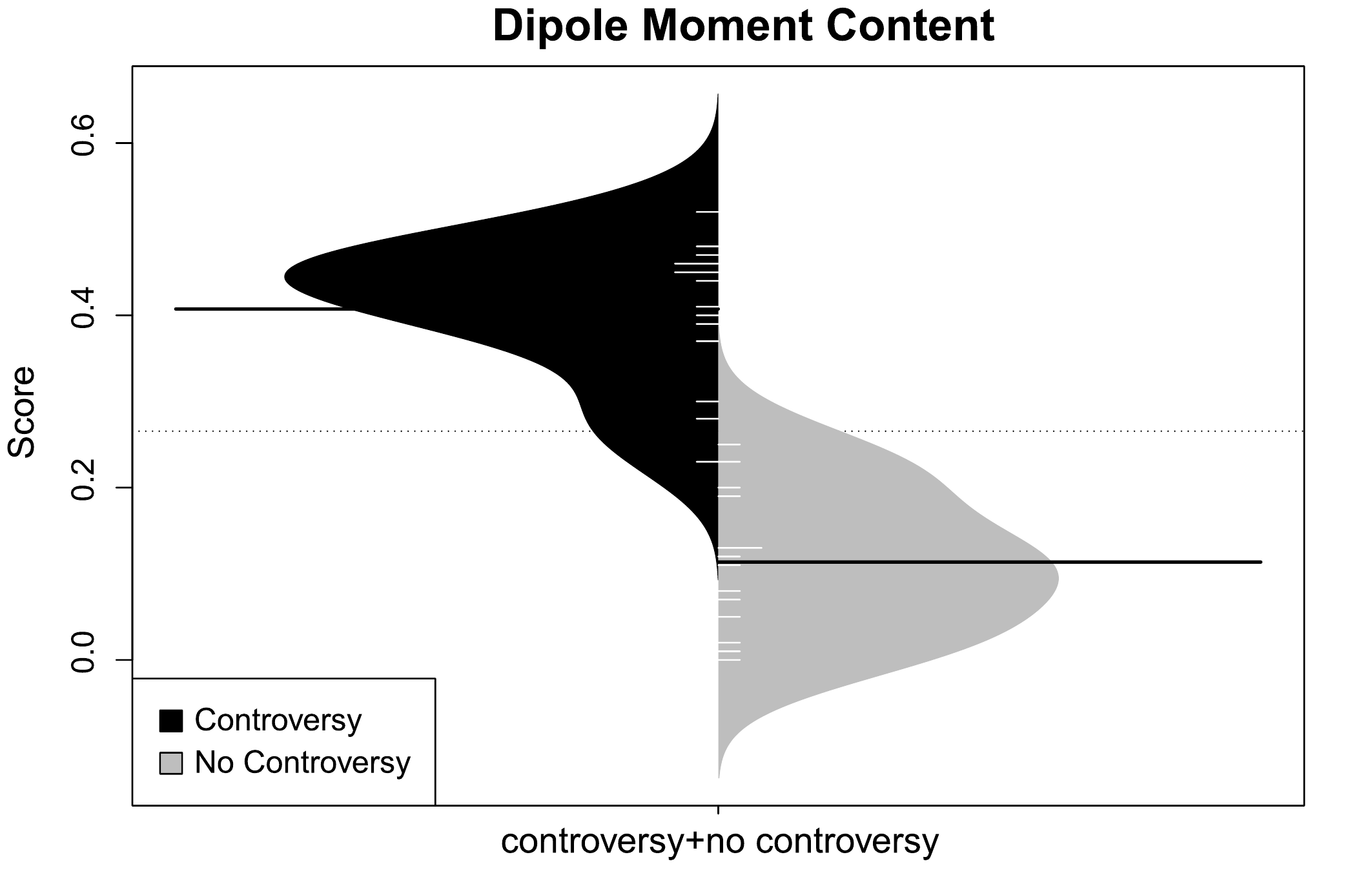}
\caption{Average controversy scores over 20 runs on datasets of 140 character per tweet}
\label{fig:scores_140_prom}
\end{subfigure}

\begin{subfigure}[t]{0.78\textheight}
    \includegraphics[width=\linewidth,height=5cm]{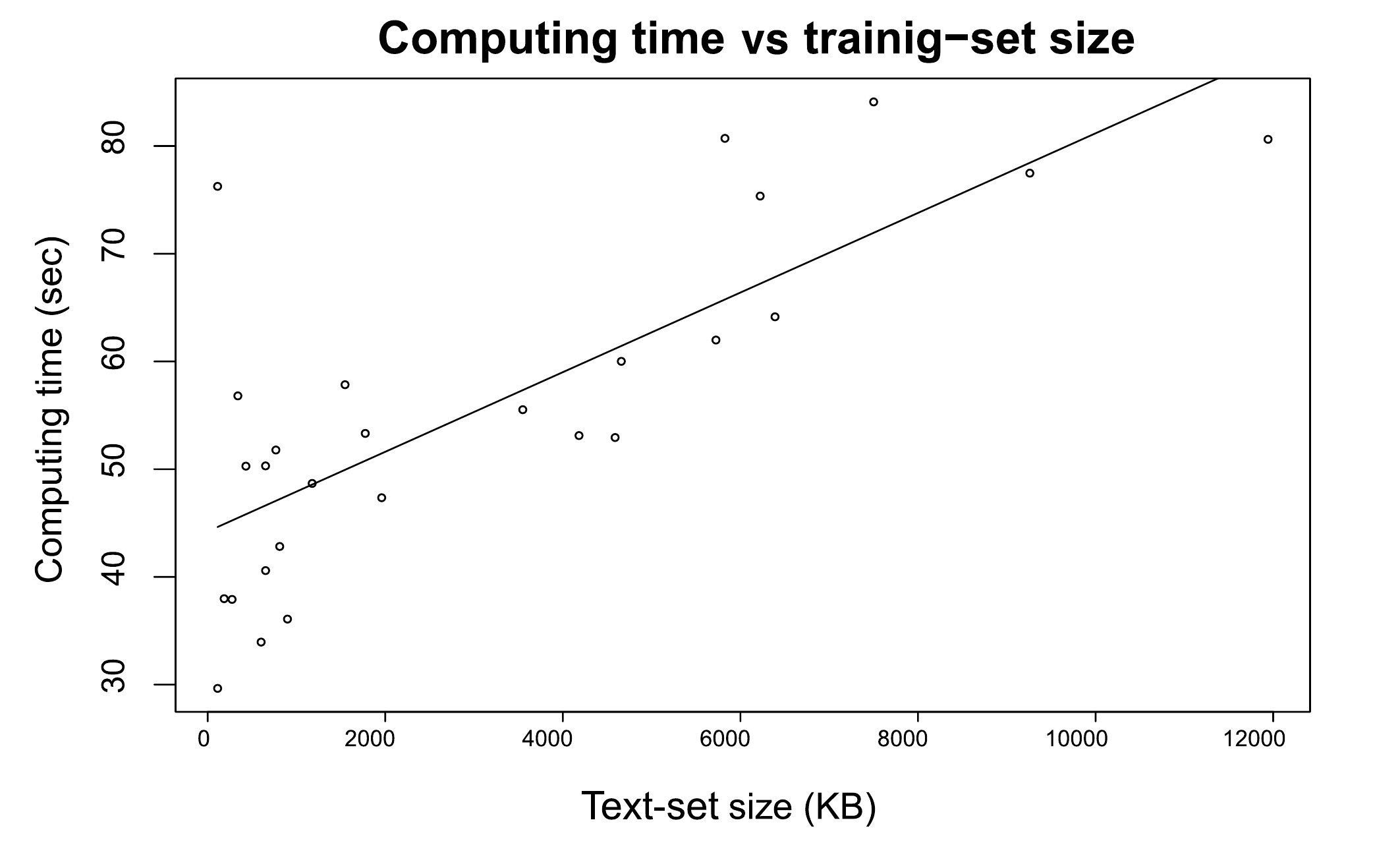}
\caption{Time text-related runs measures as a function of text-set size}
\label{fig:tiempos}
\end{subfigure}\hfill
\begin{subfigure}[t]{0.78\textheight}
    \includegraphics[width=\linewidth,height=5cm]{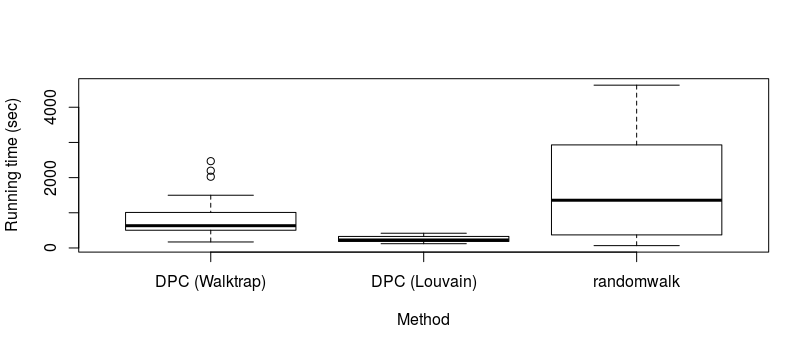}
\caption{Running time measures of our method with each cluster type and Randomwalk algorithm}
\label{fig:comparing_times}
\end{subfigure}

\caption{}

\end{figure}
\end{landscape}

Training a Fasttext model is not a deterministic process, as different runs could yield different results even using the same training set in each one. To analyze if these differences are significant, we decide to compute 20 scores for each discussion. The standard deviations among these 20 scores were low in all cases, with mean 0.01 and maximum 0.05. Consequently, we decided to report in
this paper the average between the 20 scores, in practice taking the average between 5 runs would be enough. Figure \ref{fig:scores_prom} reports the scores computed by our measure in each topic for the two cluster methods.
The beanplot shows the estimated probability density function for a measure computed on the topics, the individual observations are shown as small white lines in a one-dimensional scatter plot, and the median as a longer black line. The beanplot is divided into two groups, one for controversial topics (left/dark) and one for non-controversial ones (right/light). Hence, the black group shows the score distribution over controversial discussions and the white group over non-controversial ones. A larger separation of the two distributions indicates that the measure is better at capturing the characteristics of controversial topics, because a good separation allows to establish a threshold in the score that separates controversial and non-controversial discussions.

As we may see in the figure, the medians are well separated in both cases, with little overlapping. To better quantify this overlap we measure the sensitivity \cite{macmillan2004detection} of these predictions by measuring the area under the ROC curve (AUC ROC), obtaining a value of 0.98 for Walktrap clustering and  0.967 for Louvain (where 1 represents a perfect separation and 0.5 means that they are indistinguishable).

As Garimella et al.~\cite{garimella2018quantifying} have made their code public \footnote{https://github.com/gvrkiran/controversy-detection}, we reproduced their best method \textit{Randomwalk}\footnote{This is a measure based on random walks over the graph structure} on our datasets and measured the AUC ROC, obtaining a score of 0.935. An interesting finding was that their method had a poor performance over their own datasets. This was due to the fact (already explained in Section~\ref{sec:experiments}) that it was not possible to retrieve the complete discussions, moreover, in no case could we restore more than 50\% of the tweets. 
So we decided to remove these discussions and measure again the AUC ROC of this method, obtaining a 0.99 value. Our hypothesis is that the performance of that method was seriously hurt by the incompleteness of the data. We also tested our method on these datasets, obtaining a 0.99 AUC ROC with Walktrap and 0.989 with Louvain clustering.

We conclude that our method works better, as in practice both approaches show same performances -specially with Walktrap, but in presence of incomplete information our measure is more robust. The performance of Louvain is slightly worse but, as we mentioned in Section~\ref{sec:method}, this method is much faster. Therefore, we decided to compare the running time of our method with both clustering techniques and also with the \textit{Randomwalk} algorithm. In figure \ref{fig:comparing_times} we can see the distribution of running times of all techniques through box plots. Both versions of our method are faster than \textit{Randomwalk}, while Louvain is faster than Walktrap.

We now analyze the impact of the length of the considered text in our method. Figure \ref{fig:scores_140_prom} depicts the results of similar experiment as Figure \ref{fig:scores_prom}, but considering only 140 characters per tweet. As we may see, here the overlapping is bigger, having an AUC of 0.88. As for the impact on computing time, we observe that despite of the results of~\cite{joulin2016bag} that reported a complexity of O(h $log_{2}$(k))\footnote{Where $k$ is the number of classes and $h$ the dimension of the text representation} at training and test tasks, in practice we observed a linear growth.
%\comentario{for what? o sea, reportaron esa complejidad para qué método o algoritmo?}
We measured the running times of the training and predicting phases (the two text-related phases of our method), the resulting times are reported in figure~\ref{fig:tiempos}, which shows running time as a function of the text-size. 
We include also the best estimated function that approximate computing time as a function of text-set size.
As it may be seen, time grows almost linearly, ranging from 30 seconds for a set of 111 KB to 84 seconds for a set of 11941 KB\footnote{We compare polynomial models of degree 1 to 5 and logmodel, linear model has the lowest RMSE error training with 10-fold cross-validation.}. 
Finally, we measured running times for the whole method over each dataset with 280 characters. Times were between 170 and 2467 seconds with a mean of 842, making it in practice a reasonable amount of time. 

\section{Discussions}
\label{sec:discussion}
The task we address in this work is certainly not an easy one, and our study has some limitations, which we discuss in this section. 
Our work lead us to some conclusions regarding the overall possibility of measuring controversy through text, and what aspects need to be considered to deepen our work.

\subsection{Limitations}
As our approach to controversy is similar to that of Garimella et al.~\cite{garimella2018quantifying}, we share some of their limitations
with respect to several aspects: \textit{Evaluation} -difficulties to establish ground-truth, \textit{Multisided controversies} -controversy with more than two sides, \textit{Choice of data} - manually pick topics, and \textit{Overfitting} - small set of experiments. 
%\comentario{no entiendo lo de las 30 discussions acá}
Although we have more discussions, it is still small set from statistical point of view. 
%\comentario{alguna explicación de qué son esas cosas? si no, explicar más en general. }
\comentario{podríamos poner ``NLP-based"?}
\respuesta{podemos ponerlo, tambien text-based, content-based y mas, son todos sinónimos, creo que language es al que mas hacemos referencia}
Apart from that, our language-based approach has other limitations which we mention in the following, together with their solutions or mitigation.

\textbf{Data-size.} Training an NLP model that can predict tags with a probability greater or equal than 0.9 requires significant amount of text, therefore our method works only for ``big" discussions. Most interesting controversies are those that have consequence at a society level, in general big enough for our method.

\textbf{Multi-language discussions.} When multiple languages are participating in a discussion it is common that users tend to retweet more tweets in their own language, creating sub-communities. In this cases our model will tend to predict higher controversy scores. This is the case for example of \#germanwings,  where users tweet in English, German and Spanish and it has the highest score in no-controversial topics. However, the polarization that we tackle in this work is normally part of a society cell (a nation, a city, etc.),  and thus developed in just one language. We think that limiting the effectiveness of our analysis to single-language discussions is not a serious limitation. 

\textbf{Twitter only.} Our findings are based on datasets coming from Twitter. While this is certainly a limitation, Twitter is one of the main venues for online public discussion, and one of the few for which data is available. Hence, Twitter is a natural choice. However, Twitter's characteristic limit of 280 characters per message (140 till short time ago) is an intrinsic limitation of that network. We think that in other social networks as Facebook or Reddit our method will work even better, as having more text per user could redound on a better NLP model as we verified comparing the results with 140 and 280 characters per post.
\subsection{Conclusions}

In this article, we introduced the first large-scale systematic method for quantifying controversy in social media through content. We have shown that this method works on Spanish, English, French and Portuguese, it is context-agnostic and does not require the intervention of a domain expert.

We have compared its performance with state-of-the-art structure-based controversy measures showing that they have same performance and it is more robust. We also have shown that more text implies better performance and without significantly increasing computing time, therefore, it could be used in other contexts such as other social networks like Reddit or Facebook and we are going to test it in future works.

Training the model is not an expensive task since Fasttext has a good performance at this. However, the best performance for detecting principal communities is obtained by Walktrap. The complexity of that algorithm is O(m$n^2$)\cite{pons2005computing}, where $m$ and $n$ are the number of edges and vertices respectively. This makes this method rather expensive to compute on big networks. Nevertheless, we have shown that with Louvain the method still obtains a very similar AUC ROC (0.99 with Walktrap and 0.989 with Louvain). With incomplete information its performance gets worse but it is still good (0.96) and better than previous state of the art.

This work opens several avenues for future research. One is identifying what words, semantics/concepts or language expressions make differ one community from the other. There are various ways to do this, for instance through the word-embbedings that Fasttext returns after training \cite{joulin2016bag}. Also we could use interpretability techniques on machine learning models \cite{doshi2017towards}.
Finally, we could try other techniques for measuring controversy through text, using another NLP model as pre-trained neural network BERT~\cite{devlin2018bert} or, in a completely different approach measuring the dispersion index of the discussions word-embbedings \cite{ramponi2019vocabulary}. We are currently starting to follow this direction.

\bibliographystyle{splncs04}
\bibliography{samplepaper}
\clearpage
\begin{appendices}
\renewcommand{\thesection}{\appendixname~\Alph{section}}
% or try \arabic{section}

\section{Details on the discussions}
\label{apendice}
\begin{table*}
  \caption{Datasets statistics, the top group represent controversial topics, while the bottom one represent non-controversial ones}
  \label{tab:datasets}
  \begin{tabularx}{\textwidth}{c|c|>{\raggedright\arraybackslash}X}
    \toprule
    \multirow{1}{*}{\bfseries Hashtag/Keywords} &
    \multirow{1}{*}{\bfseries \#Tweets} & 
     
    \multirow{1}{*}{\bfseries Description and collection period}\\ 
    \midrule
    \#netanyahuspeech & 124 704  & Netanyahu at U.S. Congress,  Mar 3-5,2015 \\
    \#ukraine & 145 794  & Ukraine conflict, Feb 27–Mar 2,2015 \\
    @mauriciomacri & 108 375  & Mentions to argentian president, Jan 1–11,2018 \\
    @mauriciomacri & 120 000 & Mentions to argentian president, Mar 11-18,2018 \\
    @mauriciomacri & 147 709 & Mentions to argentian president, Mar 20-27,2018 \\
    @mauriciomacri & 309 603 & Mentions to argentian president, Apr 05–11,2018 \\
    @mauriciomacri & 254 835 & Mentions to argentian president,May 05–11,2018 \\
    Kavanaugh & 260 000 & Nomination to US supreme court, Oct 03,2018 \\
    Kavanaugh & 259 999 & Nomination to US supreme court , Oct 05,2018 \\
    Kavanaugh & 260 000 & Nomination to US supreme court, Oct 08,2018 \\
    Bolsonaro & 170 764 & Brazilian elections, Oct 27,2018 \\
    Bolsonaro & 260 000 & Brazilian elections, Oct 28,2018 \\
    Bolsonaro & 260 000 & Brazilian elections, Oct 30,2018 \\
    Lula & 250 000 & Mentions to Lula the day of Moro chats news, Jun 11-10,2019 \\
    Dilma & 209 758  & Dilma Roussef impeachment, Nov 04-06,2015 \\
    \midrule
    \#sxsw & 213 750 & SXSW conference, Mar 13–22,2015 \\
    \#1dfamheretostay & 211 979 & Last OneDirection concert, Mar 27–29,2015 \\
    \#germanwings & 199 428 & Germanwings flight crash, Mar 24–26,2015 \\
    \#mothersday & 185 964 & Mother’s day, May 8,2015 \\
    \#nepal & 249 794 & Nepal earthquake, Apr 26–29,2015 \\
    \#ultralive & 191 695 & Ultra Music Festival, Mar 18–20,2015 \\
    \#kingjacksonday & 186 263 & A GOT7 member's brithday, Mar 24–27,2019 \\
    \#Wrestlemania & 260 000 & 35th edition of Wrestlemania event, Apr 08,2019 \\
    Notredam & 200 000 & Notredam fire, Apr 16,2019 \\
    Nintendo & 203 992 & New releases by Nintendo, May 19–28,2019 \\
    Halsey & 250 000 & Halsey concert in Paris, Jun 07–08,2019 \\
    \#Bigil & 250 000& Indian actor's birthday, Jun 21–22,2019 \\
    \#VanduMuruganAJITH & 250 000 & Fans supporting actor Ajith, Jun 23,2019 \\
    Messi & 200 000 & Messi's birthday, Jun 24,2019 \\
    \#Area51 & 178 220 & Jokes about Area51, Jul 13,2019 \\
    \bottomrule
  \end{tabularx}
\end{table*}
\end{appendices}
%https://www.kpopwmx.com/2018/03/el-hastag-kingjacksonday-es-tendencia.html
%https://www.ultimahora.es/deportes/otros-deportes/2019/04/08/1071091/asi-fue-wrestlemania-noche-record-para-recordar.html
%https://www.indiatoday.in/movies/regional-cinema/story/thalapathy-63-is-now-bigil-vijay-rocks-in-two-roles-as-gangster-and-football-player-1553603-2019-06-21
%https://www.latestly.com/entertainment/south/the-war-is-between-vijay-and-ajith-fans-twitterati-use-vadivelus-pics-and-trend-vandumuruganajith-952206.html
%
% ---- Bibliography ----
%
% BibTeX users should specify bibliography style 'splncs04'.
% References will then be sorted and formatted in the correct style.
%

%

\end{document}